\documentclass[letter,longauth]{aa}     

\usepackage{graphicx}
\usepackage{txfonts}
\usepackage{lipsum}
\usepackage{subcaption}         
\usepackage{tikz}
\usepackage{pgffor}
\usepackage{lscape}             
\usepackage{placeins}           

\usepackage{array}
\usepackage{transparent}
\usepackage{xkeyval}
\usepackage{stfloats}
\makeatletter
\newlength{\sfp@hseplen}\newlength{\sfp@vseplen}
\define@cmdkey{subfigpos}[sfp@]{pos}[ul]{}
\define@cmdkey{subfigpos}[sfp@]{font}[\small]{}
\define@cmdkey{subfigpos}[sfp@]{vsep}[0.75\baselineskip]{\setlength{\sfp@vseplen}{\sfp@vsep}}
\define@cmdkey{subfigpos}[sfp@]{hsep}[3.5pt]{\setlength{\sfp@hseplen}{\sfp@hsep}}
\newcommand{\subfigimg}[4][,]{%
        \setkeys{Gin,subfigpos}{pos,font,vsep,hsep,#1}
        \setbox1=\hbox{\includegraphics{#4}}
        \ifnum\pdfstrcmp{\sfp@pos}{ul}=0
                \leavevmode\rlap{\usebox1}
                \rlap{\hspace*{\sfp@hsep}\raisebox{\dimexpr\ht1-\sfp@vsep}{\transparent{#3}{\setlength{\fboxsep}{1pt}\colorbox{white}{%
\transparent{1}\sfp@font{#2}}}%
}}
                \phantom{\usebox1}
        \else\ifnum\pdfstrcmp{\sfp@pos}{ur}=0
                \leavevmode\usebox1
                \llap{\raisebox{\dimexpr\ht1-\sfp@vsep}{\sfp@font{#2}}\hspace*{\sfp@hsep}}
        \else\ifnum\pdfstrcmp{\sfp@pos}{lr}=0
                \leavevmode\usebox1
                \llap{\raisebox{\sfp@vsep}{\sfp@font{#2}}\hspace*{\sfp@hsep}}
        \else
                \leavevmode\rlap{\usebox1}
                \rlap{\hspace*{\sfp@hseplen}\raisebox{\sfp@vsep}{\sfp@font{#2}}}
                \phantom{\usebox1}
        \fi\fi\fi
}

\usepackage{xcolor}

\newcommand{\KM}[1]{\textcolor{black}{#1}}

\begin{document}

   \title{Unmasking (44) Nysa: Evidence for a trilobate structure}


%
%
%

   \author{K. Minker\inst{1}\corrauth{kminker@lowell.edu}        
        \and A. Berdeu\inst{2}\corrauth{Anthony.Berdeu@obspm.fr}
        \and G. Li Causi\inst{3}
        \and J. Hanu\v{s}\inst{4}
        \and M. Marsset\inst{5}
        \and A. Conrad\inst{6}
        \and F. Pedichini\inst{3}
        \and S. Antoniucci\inst{3}
        \and P. Vaccari\inst{3}
        \and N. Moskovitz\inst{1}
        \and B. Carry\inst{7}
        \and T. Polakis\inst{8}
        \and M. Ferrais\inst{9}
        \and E. Jehin\inst{10}
        \and R. Bonamico\inst{11}
        \and R. Hicks\inst{9}
        \and D. Smith\inst{9}
        \and M.A. Miftah\inst{10,12}
        \and E. Marini\inst{3}
        }

   \institute{Lowell Observatory, 1400 W. Mars Hill Rd., Flagstaff, AZ, USA
   \and European Southern Observatory (ESO), Santiago, Chile 
   \and INAF-Osservatorio Astronomico di Roma, Via Frascati, 33, Monteporzio Catone, 00078, Italy
   \and Charles University, Faculty of Mathematics and Physics, Institute of Astronomy, V Hole\v{s}ovi\v{c}k\'ach 2, 180\,00 Prague, Czech Republic
   \and European Southern Observatory (ESO), Karl-Schwarzschild-Strasse 2, 85748 Garching bei München, Germany
   \and Large Binocular Telescope Observatory, Tucson, AZ, USA
   \and Université Côte d'Azur,  Observatoire de la Côte d'Azur, Lagrange, CNRS, Nice, France
   \and Command Module Observatory, Tempe, AZ, USA
   \and Florida Space Institute, University of Central Florida, 12354 Research Parkway, Orlando, FL 32828, USA
   \and Space Sciences, Technologies \& Astrophysics Research (STAR) Institute, University of Liège, Liège, Belgium
   \and BSA Osservatorio (K76), Strada Collarelle 53, 12038 Savigliano, Cuneo, Italy
   \and Cadi Ayyad University (UCA), Oukaimeden Observatory (OUCA), Faculté des Sciences Semlalia (FSSM), High Energy Physics, Astrophysics and Geoscience Laboratory (LPHEAG), Marrakech, Morocco.}

   \date{Received September 30, 20XX}

 
  \abstract
   {(44) Nysa is one of the largest known E-type asteroids. Light curve inversion models have indicated that it may have an elongated shape commonly found in large binary systems.}
   {We aimed to identify the morphology and presence of possible satellites orbiting  Nysa.} 
   {We observed Nysa with the visible-wavelengths adaptive-optics instruments SHARK-VIS and SPHERE/ZIMPOL in order to image the object with the highest possible spatial resolution, and visually identified surface features on the object. A shape model was constructed from these images and photometric light curves using the ADAM code.}
   {Imaging revealed a highly unusual morphology with multiple distinct surface features, including two distinct valleys most easily interpreted as colli. A faint satellite was identified in the deconvolution residuals in multiple datasets.}
   {We determine that (44) Nysa is a binary system most likely consisting of a contact-trinary or highly-irregular, coherent primary object and a small satellite.}


   \maketitle
\nolinenumbers

\section{Introduction}
\KM{Asteroid} (44) Nysa was named to honor the mythical land of Nysa where the god Dionysus (also known as Bacchus) was raised by nymphs \citep{2012dmpn.book.....S}. Dionysus is associated with masks and theater, known for concealing his true nature. The naming of Nysa for the land in which this god of illusion was raised is quite fitting, as asteroid Nysa has long managed to evade attempts to identify its true nature, leaving only vague clues to the object's structure. Nysa was observed with the Hubble Space Telescope's Fine Guidance Sensor, by radar from Arecibo Observatory, and by stellar occultation, which collectively identified a potential concavity or bilobate structure, but no concrete determination of the asteroid's morphology was achieved through these efforts \citep{2002ESASP.500..517T, 2008Icar..195..220S, 2019pds..rept.....H}. A light curve inversion model for Nysa's shape suggested an elongated morphology typical of binary systems with large primaries \citep{2002A&A...383L..19K, 2021A&A...650A.129C, 2025A&A...701A..42M}.

Nysa is one of the largest known E-type asteroids \citep{2022A&A...665A..26M}, and the brightest as the only E-type with $H<7$. This rare taxonomic type is associated with enstatite-rich meteorites including aubrites and enstatite chondrites \citep{1977GeCoA..41.1759Z,2010ChEG...70..295K} and is thought to be formed from the same feedstocks as the terrestrial planets. \KM{As defined by \citet{2004JGRE..109.2001C}}, Nysa-like E-types form a subcategory of this compositional group which exhibit additional orthopyroxene features  These objects do not provide a clear spectroscopic match to any known meteorites, although this complex problem may be biased by grain size effects and other influences \citep[e.g.,][]{Cantillo_2026}. The Nysa-like \KM{spectroscopic} profile has been detected in small near-Earth asteroids \citep[NEAs, e.g.][]{2016AJ....152..162R}, including most recently the future Hayabusa2\# target 1998 $KY_{26}$ \citep{2026A&A...709A.105T}.


Nysa's large size and high albedo \citep[reports range from 0.41-0.55, see][and references within]{2023A&A...671A.151B} position \KM{the asteroid} as a particularly attractive observational target. As a result, it was one of the first asteroids to have a measured rotation period by photometric light curve 
\citep[e.g.][]{Bianchi1920}. Nysa shows both a higher polarization ratio ($\mu=0.50\pm0.02$) and radar albedo ($0.19\pm0.06$) compared to the majority of main-belt asteroids, suggesting a high surface bulk density \citep{2008Icar..195..220S}.

\KM{Recent} observations by SHARK-VIS on the Large Binocular Telescope (LBT) managed to reveal \KM{the} highly unusual morphology \KM{of (44) Nysa} \citep{2026CBET.5664....1M}. These observations and their interpretation are discussed in the remainder of this letter.

\section{Observations}

\subsection{Adaptive optics imaging}\label{sec:img}

We observed Nysa with SHARK-VIS \citep{2022SPIE12185E..6QP, 2024SPIE13097E..0AP} on UT 2026-02-15 and UT 2026-03-21 (program AZ-2026A-004 and DDT), and with VLT/SPHERE/ZIMPOL \citep{2019A&A...631A.155B, 2018A&A...619A...9S} (program 116.2AQK.001) on ten epochs between 2026-03-09 and 2026-04-05.

SHARK-VIS acquired images at the F/15 Nasmyth focal plane of the LBT after atmospheric correction by the SOUL adaptive optics system \citep{2016SPIE.9909E..3VP}. \KM{On} UT 2026-02-15 we acquired DIT=25\,ms exposure frame sequences with the $R_{Bessel}$ filter 
\KM{between} UT 04:20 and UT 07:44 \KM{with seeing conditions stable at 0.8". On} 2026-03-21 we acquired DIT=80\,ms exposures \KM{alternating between $R_{Bessel}$ and $V_{Bessel}$ filters between UT 02:34 and UT 05:57 with seeing variable between 0.8-1.1"}. 
\KM{Both acquisitions used} a field stop with a diameter of 1.2". The plate scale was 6.43\,mas/px on the first date, while on the second date \KM{an} internal magnification lens \KM{was used} which yielded an effective plate scale \KM{of $\approx$}4.45\,mas/px to better sample the image.  Additional details can be found in Table~\ref{tab:shark}.

The ZIMPOL dataset consists of 10 identical sequences of images in the \verb|N_R|, V, Cnt748 and Cnt820 filters, recorded as two sets of two simultaneously imaged filters (\verb|N_R| \& V, Cnt748 \& Cnt820). The 10 observations were timed to cover the full rotation period and phase angles ranging from 21.6--$27.1^o$. Each set of images included six 15\,s exposures, for a total of 90\,s exposure time per filter.  Seeing ranged between 0.3-0.8" for different epochs of observation. The timing and orbital geometry of these observations can be found in Table~\ref{tab:sphere}.

The two imaging datasets were then processed with dark, flat, and cosmic ray correction, lucky imaging (in the case of the LBT observations), and blind deconvolution algorithms to maximize image quality and improve the visibility of surface features across the primary object. The techniques used for the reduction of these images can be found in App.~\ref{sec:app_AO}. The final processed video can be found in Visualization 1 (online video).

\subsection{Photometric light curves}\label{sec:lc}

Photometric light curve observations were used to supplement the imaging dataset. Light curves were acquired at the 0.6\,m TRAPPIST-North and -South telescopes \citep{2011Msngr.145....2J}, BSA Observatory, the 0.5\,m telescope at Robinson Observatory, and the 0.32\,m telescope at Command Module Observatory. Some light curves were collected in multiple filters to search for color variation across Nysa's rotation period.
Archival observations were collected from ALCDEF\footnote{\url{https://alcdef.org/}} \citep{2016MPBu...43...26W}. Most photometric light curve observations depict only small deviations from a sinusoid, although some present a singular flat minimum, as is typical of contact binaries and previously reported by \citet{1985A&A...144..355C}. A summary of the photometric dataset can be found in Table~\ref{tab:lc}, and is illustrated alongside synthetic lightcurves produced from the shape model discussed in Sec. \ref{sec:shape}.


\section{Shape model}
\label{sec:shape}

A three-dimensional shape model of Nysa was reconstructed with the All-Data Asteroid Modeling algorithm \citep[ADAM;][]{2015A&A...576A...8V}, combining the disk-resolved images described in Sec.~\ref{sec:img} with the photometric lightcurves in Sec.~\ref{sec:lc}.

Nysa’s rotation period, close to 6.42~h, has been known since the 1920s and has subsequently been confirmed and refined by several studies \citep[e.g.,][]{Bianchi1920, Groeneveld1954b}. Previous lightcurve-based spin-state analyses generally reported the usual mirror pole ambiguity, with solutions near ecliptic coordinates ($\lambda,\beta)\approx(100^{\circ},60^{\circ}$) and ($280^{\circ},60^{\circ}$) \citep[e.g.,][]{Magnusson1983,Taylor1983}. By contrast, the convex lightcurve inversion model of \citet{2002A&A...383L..19K} yielded a unique pole solution near ($99^{\circ},58^{\circ}$). Our disk-resolved images clearly confirm this solution.

We followed the ADAM modeling strategy previously applied in several SPHERE-based studies, for example for asteroids (16)~Psyche \citep{Viikinkoski2018}, (7)~Iris \citep{Hanus2018d}, and (31)~Euphrosyne \citep{Yang2020a}, as well as in the survey summary by \citet{Vernazza2021}. The convex spin-state solution was adopted as the initial guess, ensuring convergence to the correct minimum. This solution, derived from the full set of available lightcurves, is close to that reported by \citet{2002A&A...383L..19K}. We first computed a low-resolution model with 400 vertices and 800 facets, and then gradually increased the surface resolution while monitoring convergence and the reproducibility of topographic features by adjusting the regularization terms. Our final model, with 1600 vertices and 3200 facets, provides a good fit to both the observed lightcurves and all disk-resolved images. Further increasing the resolution does not improve the fit. The solution is stable with respect to changes in surface resolution, data weighting, and reasonable variations of the regularization parameters. \KM{Our ADAM configuration includes a regularization term that forces the calculated center of mass of the shape model to be as close as possible to the center of the coordinate system. Therefore, no information about the internal mass distribution can be derived from the model. }

We also tested the consistency of the model against three stellar occultations with at least four chords (2007-01, 2017-09, 2017-10), reduced using the \texttt{Occult} software package\footnote{\url{http://www.lunar-occultations.com/iota/occult4.htm}}. Because the number of chords is limited and the sampled silhouettes are sparse, these occultations were not included directly in the shape reconstruction. Nevertheless, they are consistent with our preferred model and thus provide an independent validation of the solution (Fig.~\ref{fig:occ}).

The derived spin-state parameters determine spin-pole coordinates of ($\lambda$, $\beta$)=(102.7$\pm$0.5$^{\circ}$, 55.3$\pm$0.5$^{\circ}$), a rotation period of $P$=6.4214175$\pm$0.0000005)~h, and a volume-equivalent diameter of $D$=75$^{+3}_{-1}$~km.
A comparison between the shape model and the AO images as well as the fits to the lightcurves can be found in Fig. \ref{fig:nysa_lcs_1}.

\begin{figure}
    \centering
    \includegraphics[width=0.8\linewidth]{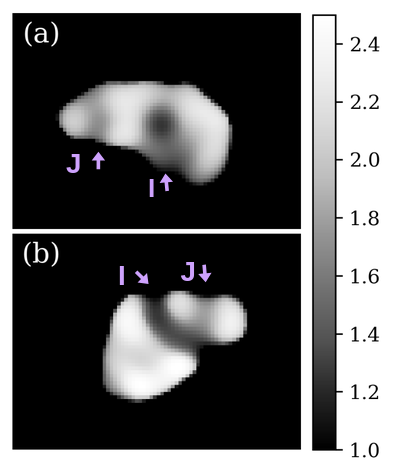}
    \caption{2026-03-21 SHARK-VIS observations of Nysa, color scale shows relative brightness and is tuned to enhance the visibility of shadows, indicative of surface indentations or valleys in the \KM{sky-East (upper panel) and -West (lower panel)} facing geometries. Two clear shadowed regions can be seen, one at the object's center (I) and another close to its narrowest point (J). Shadows are apparent in multiple geometries, appear to fully encircle the asteroid, and are bounded by edge indentations. This suggests that both are colli connecting a total of three lobes. Additional features are annotated in Fig. \ref{fig:Nysacraters}. }
    \label{fig:contactriple}
\end{figure}

\section{Surface features}

This dataset revealed a number of surface features across Nysa. \KM{Features were divided into two categories}; indentations on the edges of the object and \KM{valleys} or shadows on the center of the object. All features reported here are consistent with the rotation of the object and most are detected on multiple nights.

A map of identified surface features can be found in figure \ref{fig:Nysacraters}. We identified eleven notable features in the imaging dataset; eight edge indentations and two valleys. The valleys are most visible in higher phase-angle observations from the March dataset, this can be seen in figure \ref{fig:contactriple}. The valleys appear when the asteroid is in an \KM{on-sky} East-facing or West-facing geometry (see Fig. \ref{fig:contactriple}), suggesting that they transect the full circumference of the asteroid along its narrowest axis, and join edge indentations on either side. We interpret these valleys as colli \KM{(necks)} joining contact-fragments. The E-I-F (see Fig.~\ref{fig:contactriple}, Fig.~\ref{fig:Nysacraters} and Tab.~\ref{tab:craters}) collum joins the two main lobes of Nysa, while the E-J-A collum separates the two halves of the smaller lobe. In this interpretation and taking all remaining edge indentations to be craters, we identify three lobes, two colli, and five craters.

\citet{2003A&A...408..379T} identified a U-B color variation across Nysa's rotation. We observed no color variation through multi-filter light curves or in between lobes in the SPHERE dataset at greater than a 5\% level, well below the variation of 0.2\,mag reported by \citet{2003A&A...408..379T}.

\section{S/2026 (44) 1}
\label{sec:satellite}
A small satellite was identified in LBT observations from both 2026-02-15 and 2026-03-21 \citep{2026CBET.5693....1B}. The satellite was detected both (I) in the residuals of the blind deconvolutions that model and subtract the bright halo around Nysa and (II) from \KM{an independent} Principal Component Analysis (PCA) - Angular Differential Imaging (ADI) analysis 
inspired from exoplanet detection (see Secs.~\ref{sec:app_blind} and ~\ref{sec:app_PCA} and Figs. ~\ref{fig:moonPCA} and ~\ref{fig:deconv}).

\begin{figure}
    \centering
    \includegraphics[width=0.75\linewidth]{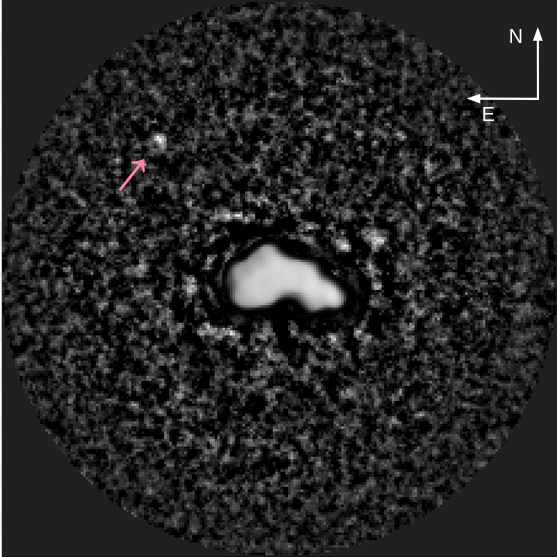}
    \caption{
    \KM{Residuals of R-band images of the Nysa system on 2026-02-15 after processing with the blind deconvolution method of Appendix~\ref{sec:app_blind}. The position of the satellite is marked with a pink arrow. To improve the S/N, the deconvolution residuals were averaged with a a $\pm5$ frame sliding window. To provide scale, the primary has been superimposed.} An extended video version of this figure and the same processing for the 2026-03-21 dataset is available online.}
    \label{fig:moonsuperimposed}
\end{figure}

The identified point source was clearly moving on sky with the trajectories depicted in Fig.~\ref{fig:moonPCA} while remaining present in a Nysa-centered frame for over three hours in each epoch. This removes the possibility of contamination from a \KM{background object}.
The satellite was detected at a projected separation of 180\,km and 170\,km in the February and March observations respectively. \KM{A $\Delta_{mag}\approx9$ was measured between components.}
Assuming consistent albedo between components, this provides a satellite diameter of $D_s=1\pm0.5$\,km. Orbital scale with respect to the size of the primary is demonstrated in Fig. \ref{fig:moonsuperimposed}. Additional study of this satellite will be reported in future work.

\section{Discussion}\label{sec:discussion}

\subsection{Atypical morphology}\label{sec:shapediscussion}

Several critical constraints restrict possible formation mechanisms for Nysa's abnormal morphology. Namely:
\begin{itemize}
    \item The mechanism should not require the ejection of a large amount of material in Nysa's current position.
    Although Nysa is embedded in the expansive Nysa-Polana complex \citep{2001Icar..152..225C}, only a small fraction of material in this complex exhibits an albedo high enough to indicate enstatite material \citep{2024A&A...689A.183B, 2026A&A...709A.105T}. Neither the total quantity nor size of individual fragments of high-albedo material in the Nysa-Polana complex is sufficient to indicate a recent dramatic re-shaping of the object. 
    \item Dispersed Nysa-like E-type asteroids are observed scattered throughout the main-belt \citep{2004JGRE..109.2001C}, as well as in NEA populations \citep{2016AJ....152..162R, 2026A&A...709A.105T}, so a formation event that occurs prior to implantation in the main-belt could explain the lack of a dynamical center for the \KM{Nysa-like} material that would be expected from a recent formation.
    \item Ideally, the mechanism should provide a possibility of low-velocity mergers/re-accumulation of material from a single parent body. Nysa displays at least two deep valleys which circumscribe the object. Our preferred interpretation of these features is that they are colli connecting three distinct lobes.
    Nysa's  surface composition, representing only a fraction of already rare E-type asteroids \citep[the "Nysa-like" subset][]{2004JGRE..109.2001C}, \KM{shows no color or albedo difference between} lobes of the object \KM{in the observed filters}. This suggests that all surface material on Nysa came from the same parent body, and by extension that all of the lobes should be sourced from said parent body.
\end{itemize}

We present possible scenarios for the origin of Nysa's highly unusual shape in the following subsections. 

\subsubsection{Battering}

 The simplest explanation for Nysa's unusual morphology is that \KM{Nysa} is a heavily battered object that originated from a singular object which has been heavily deformed due to substantial cratering. If this interpretation is correct, Nysa must have a sturdy interior to survive these nearly catastrophic collisions. 
 The cratering events may not be ongoing, as very little enstatite material is observed in Nysa's orbital neighborhood. 
 The detection of a very high density ($\rho\gtrapprox5$\,g.cm$^{-3}$) could support, but not necessarily confirm, this possibility by suggesting the presence of a rigid, iron-rich core which could support this distorted morphology. This mechanism does not adequately explain the presence of the two valleys that transect Nysa's surface along its narrowest axis, and is therefore not our preferred explanation. Alternatively, it could be possible that Nysa is a single, coherent planetary shard whose bizarre shape originates from impact-based reshaping along pre-existing fractures. In this scenario, the battering would be responsible solely for the smaller concavities on the object. A lower, aubrite-like ($3-4$\,g.cm$^{-3}$) density would be expected.

\subsubsection{Contact trinary}
\KM{A second explanation is that} Nysa is a composite of several components in a contact trinary (or more) configuration. This formation possibility is supported by the presence of multiple circum-asteroid valleys across Nysa's narrowest axis, \KM{typically}
interpreted as a collum joining two distinct components as a contact binary \citep[\KM{e.g. Toutatis,}][]{2014GeoRL..41..328Z}. In some cases, a dramatic indentation can be found at the collum, with little evidence of relaxation over time, as is the case for (486958) Arrokoth \citep[][]{2019Sci...364.9771S}. Nysa is notably larger than any of these objects, and the central concavity is much more extreme than any previously observed. 
\KM{Therefore, this interpretation indicates that} Nysa would need to be sturdy enough to maintain its seemingly-unstable structure through the small impacts over the object's lifetime, and possibly implantation to the main-belt. Alternatively, it could be possible that the three components of Nysa are individually sturdy, but their specific configuration has varied over time as the system has been disrupted. While most well-studied asteroids are fairly low-cohesion, if Nysa and its sub-components possess higher cohesion, it may be easier for Nysa to retain its unusual shape over time, as this material would be more likely to retain irregular shard-like topography after impacts \citep[e.g.,][]{2010Sci...327..190K}. 
Further analysis of the long-term stability of the system will be presented in future work.

\subsection{Formation}

Various proposed mechanisms could cause the observed structure. 
However, we note that Nysa does indeed match the elongated, lopsided, fast rotating profile associated with large binary primaries \citep[][]{2021A&A...650A.129C, minke2025}. It is possible that the shape originates from a "hit-and-run collision" as described by \citet{2006Natur.439..155A}. This process has been proposed to highly deform differentiated asteroids as tidal stresses effect the lower density mantle layer more strongly than the dense core, and are capable of stripping an object into similarly-sized ("string-of-pearls") fragments \citep{2006Natur.439..155A}. Nysa could be an accumulation of such fragments shortly after the initial deformation reassembled via low-velocity merger \citep[e.g.,][]{2025NatCo..1611033R}, explaining the \KM{apparent surface homogeneity}.

The possibility of reaccumulation following a giant impact to a proto-Nysa cannot be excluded. However, it seems unlikely that such a re-accumulation could occur in-situ as an impact capable of creating three 60-20\,km sized objects which then reaccumulate should also produce additional debris, which would be observed as a large, enstatite-rich dynamical family of which Nysa is the largest member. Such a family has not been detected \citep{2015Icar..252..199D}.

\section{Conclusions}
By observing (44) Nysa with optical-wavelength adaptive optics imaging, 
we were able to \KM{identify} both a contact-trinary morphology and a small satellite. The origin of this structure is not yet clear, but it may have originated prior to Nysa's implantation to the main belt due to the lack of enstatite material in the dynamical space surrounding Nysa.
Ongoing work on the satellite dynamics will provide a refined density constraint, and together with modeling of the long-term stability and formation of the Nysa system will reveal additional clues to the objects internal structure and origin.


%

\bibliographystyle{bibtex/aa}
\bibliography{bibtex/bib,bibtex/bib2}

\begin{appendix}
\nolinenumbers





\begin{acknowledgements}
      K.M. acknowledges support through the Percival Lowell Postdoctoral Fellowship which is funded in part through generous donations of the Percival Lowell Society. J.H. was supported by GACR grant no. 25-16789S of the Czech Science Foundation.
      We wish to thank Joseph Shields and the ESO DDT selection committee for the allocation of director's discretionary time to this project, Johan Olofsson and Jennifer Power for the observing support, Richard Jerousek and Christopher Duffey for their contributions to the Robinson Observatory.
      
      \KM{The LBT is an international collaboration among institutions in the United States and Europe. At the time data were acquired for this research, LBT Corporation Members were the University of Arizona on behalf of the Arizona Board of Regents; Istituto Nazionale di Astrofisica, Italy; and The Ohio State University, representing The Ohio State University, University of Notre Dame, University of Minnesota, and University of Virginia.  This research used the facilities of the Italian Center for Astronomical Archives (IA2) operated by INAF at the Astronomical Observatory of Trieste.  Observations have benefited from the use of ALTA Center (alta.arcetri.inaf.it) forecasts performed with the Astro-Meso-Nh model. Initialization data of the ALTA automatic forecast system come from the General Circulation Model (HRES) of the European Centre for Medium Range Weather Forecasts.  TRAPPIST is funded by the Belgian F.R.S.-FNRS under grant PDR T.0120.21, and TRAPPISTNorth is funded by the University of Liège in collaboration with Cadi Ayyad University of Marrakech. EJ is Director of Research at the Belgian National Fund for Scientific Research (F.R.S.-FNRS). }
      Thank you to E. Asphaug, A. Thirouin, and M. Goldberg for the constructive discussion.
\end{acknowledgements}

\section{Adaptive optics data reduction}

\label{sec:app_AO}

\subsection{Raw data reduction}
\label{sec:app_data}

We processed the SHARK-VIS frames with the custom DReAMS (Data Reduction, Analysis, and Modeling for SHARK-VIS) pipeline (Li Causi et al., in prep). The raw frames were reduced through various steps to correct for detector signatures: (I) subtraction of a cleaned average dark image, separately acquired at the end of the observing night; (II) correction of the frame-by-frame variable bias pattern, mainly consisting of non-static column-by-column (and less intense row-by-row) bias variations, which we measured on each frame as the column and row averages, respectively, of the pixels outside the field stop; (III) cosmic rays flagging; (IV) division by a flat field map, interpolated from a flat field ramp previously acquired in the instrument calibration phase.

The SPHERE observations were obtained in imaging mode and processed using the standard ZIMPOL imaging pipeline through the ESO Data Processing System (EDPS; \citealt{2024A&A...681A..93F}). The pipeline performs standard calibrations including flat-field correction, bad-pixel identification, 
and image orientation.
To enable high-precision polarimetric observations, ZIMPOL employs a masked CCD synchronized with a fast polarization modulator. In non-polarimetric imaging, this results in one row out of two not being used. To recover a uniform pixel sampling, the reduction pipeline reconstructs the final image by interpolating across these masked rows. This reconstruction process can leave low-level striping artifacts visible in the deconvolved images \KM{(see Fig. \ref{fig:shape1})}.

\subsection{Pre-processing SHARK-VIS short exposures}

The idea behind the SHARK-VIS short exposures was to freeze the turbulence and the residual jitter from the AO system to increase image quality. Nonetheless, taken individually, they are too noisy and need to be adequately stacked. We implemented two different strategies depending on the science need.

\subsubsection{Nysa's shape: lucky imaging and super-resolution}
\label{sec:app_lucky}

For Nysa's imaging and shape recovery, the sharpest possible images are needed. As pre-processing for the blind deconvolution of Sect.~\ref{sec:app_blind}, we performed lucky imaging. We split the exposures in batches of 2~minutes, corresponding to 4800 (resp. 1500) frames for the dataset of UT 2026-02-15 (resp. UT 2026-03-21) to limit the motion blur due to Nysa's proper rotation and the evolution of the parallactic angle. The frames were sorted by their sharpness computed via total-variation~\citep{Charbonnier:1997_TV} and then co-aligned on Nysa's photocenter. Then, this alignment was iteratively refined in the resulting stacked image, iterating thrice. The sub-pixel alignment and stacking was done via linear interpolation.

We kept the best 2.5\,\% of frames for the UT 2026-02-15 night. To compensate the loss of one magnitude due to Nysa's farther distance and the optical magnification while keeping a similar S/N, we kept the 20\,\% best ones for the UT 2026-03-21 night.

Finally, for the UT 2026-02-15 dataset, the interpolation grid was defined with a resolution twice as high as the pixel pitch to perform numerical super-resolution via the dithering induced by the residual AO jitter. Figure~\ref{fig:deconv}a gives an example of a stacked batch of 2~minutes, while Fig.~\ref{fig:deconv}b shows its corresponding lucky imaging reduction with the factor 2 resolution enhancement. As the UT 2026-03-21 dataset was obtained with optical magnification, we did not super-resolve the reduced frames.

\subsubsection{Moon detection: co-alignment of all frames}
\label{sec:app_co_align}

For companion detection, the highest possible S/N is needed. As pre-processing for the PCA-ADI analysis described in Sect. \ref{sec:app_PCA}, we used the full dataset regardless of image quality, in order to increase the signal to noise ratio. Individual frames were aligned on Nysa's photocenter. Their sub-pixel alignment was performed with a Fast Fourier Transform translation. 

\begin{figure}
    \centering
    \includegraphics[width=0.75\linewidth]{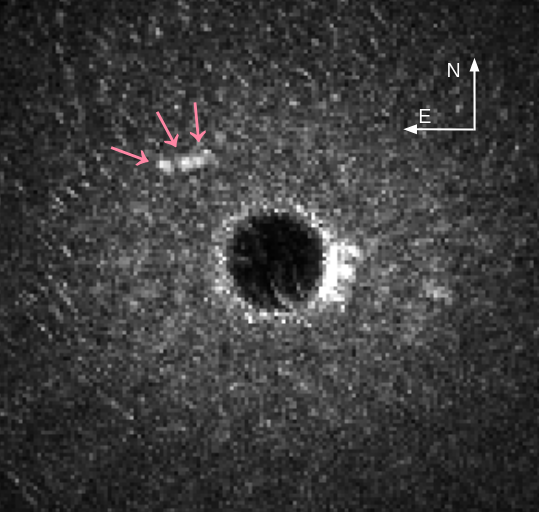}
    \caption{Maximum pixel stack of PCA-ADI treated R-band images of the Nysa system on 2026-02-15, three detected positions of the moon on this night are superimposed and marked with pink arrows. }
    \label{fig:moonPCA}
\end{figure}

\subsection{Blind deconvolution algorithm}
\label{sec:app_blind}

\begin{figure}[h!]
    \centering
	\newcommand{\ColumnWidth}{4.45cm}
	\newcommand{\ColumnGap}{\hspace{1.5pt}}
	\newcommand{\subfigColor}{white}
	\newcommand{\fontfig}[1]{\scriptsize$\!$\color{#1}\textbf}

	\begin{tabular}{
		@{}
		>{\centering\arraybackslash}m{\ColumnWidth}
		@{\ColumnGap}
		>{\centering\arraybackslash}m{\ColumnWidth}
		@{\ColumnGap}
		>{\centering\arraybackslash}m{\ColumnWidth}
		@{}
		}
		\subfigimg[width=\linewidth,pos=ul,font=\fontfig{\subfigColor}]{(a)}{0.0}{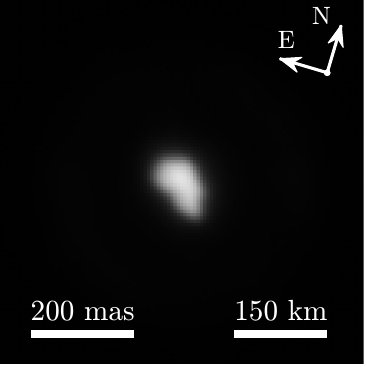} 
		&
		\subfigimg[width=\linewidth,pos=ul,font=\fontfig{\subfigColor}]{(b)}{0.0}{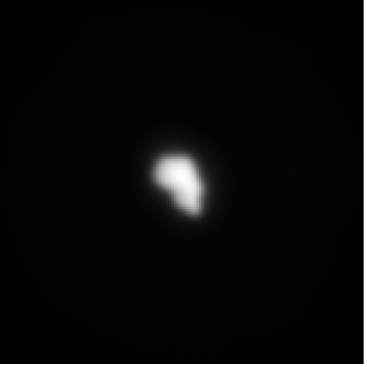} 
		\\[-2pt]
		\subfigimg[width=\linewidth,pos=ul,font=\fontfig{\subfigColor}]{}{0.0}{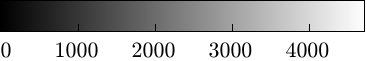} 
		&
		\subfigimg[width=\linewidth,pos=ul,font=\fontfig{\subfigColor}]{}{0.0}{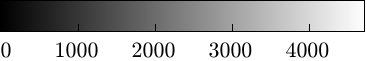} 
		\\[2pt]
		\subfigimg[width=\linewidth,pos=ul,font=\fontfig{\subfigColor}]{(c)}{0.0}{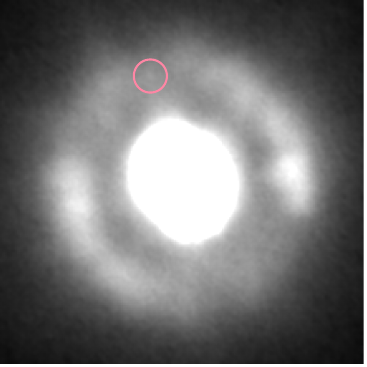} 
		&
		\subfigimg[width=\linewidth,pos=ul,font=\fontfig{\subfigColor}]{(d)}{0.0}{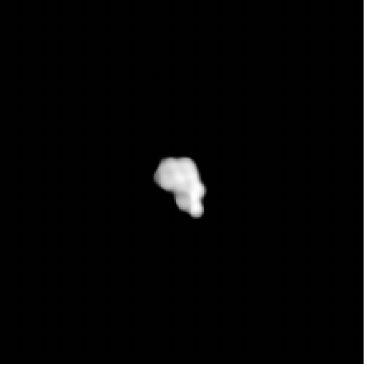} 
		\\[-2pt]
		\subfigimg[width=\linewidth,pos=ul,font=\fontfig{\subfigColor}]{}{0.0}{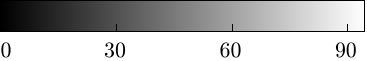} 
		&
		\subfigimg[width=\linewidth,pos=ul,font=\fontfig{\subfigColor}]{}{0.0}{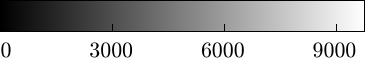} 
		\\[2pt]
		\subfigimg[width=\linewidth,pos=ul,font=\fontfig{\subfigColor}]{(e)}{0.0}{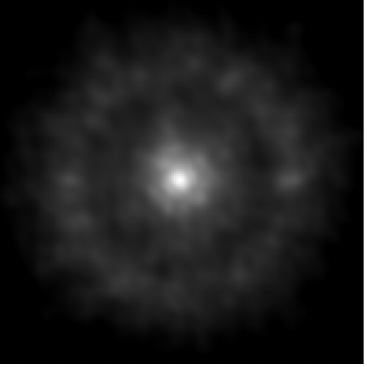} 
		&
		\subfigimg[width=\linewidth,pos=ul,font=\fontfig{\subfigColor}]{(f)}{0.0}{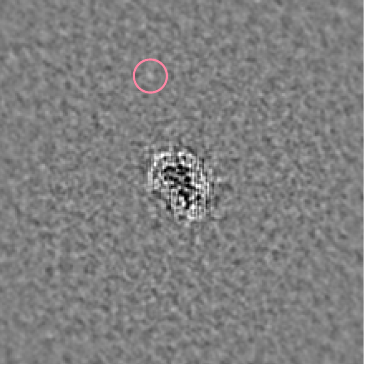} 
		\\[-2pt]
		\subfigimg[width=\linewidth,pos=ul,font=\fontfig{\subfigColor}]{}{0.0}{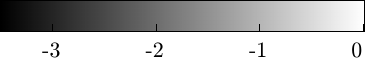} 
		&
		\subfigimg[width=\linewidth,pos=ul,font=\fontfig{\subfigColor}]{}{0.0}{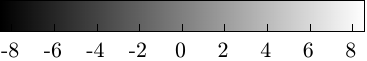} 				        
	\end{tabular}
    \caption{Applying the blind deconvolution algorithm on a SHARK-VIS 2 min batch of the UT 2026-02-15 night. Panel~a: stack of batch. Panel~b: lucky imaging and super-resolution of the 2 min batch, keeping the 2.5\,\% best frames. Panel~c: saturated view of Panel~b. Panel~d: deconvolved image. Panel~e: reconstructed PSF. Panel~f: deconvolution residuals.
    Except for the PSF where the color scale is in log scale normalized to the PSF peak, all color scales are in ADU. The colored circle highlights Nysa's moon S/2026 (44) 1. The image orientation is in the pupil frame.}
    \label{fig:deconv}
\end{figure}

To deconvolve the different datasets (the long exposure ZIMPOL data, see Sect.~\ref{sec:app_data} and the SHARK-VIS stacked frames after lucky imaging, see Sect.~\ref{sec:app_lucky}), we followed the method of \citet{2024A&A...688A..18B}, upgraded with a better segmentation of the asteroid described by \citet{Berdeu:AO4ELT8}. The method performs a blind deconvolution of the images, jointly recovering (I) the de-blurred image of the asteroid, see Fig.~\ref{fig:deconv}d and, (II) the AO PSF in its full complexity, see Fig.~\ref{fig:deconv}e. Doing so allows \KM{for} modeling the bright halo produced by the AO residuals and AO cutoff frequency visible in Fig.~\ref{fig:deconv}c. Removing its model from the data highlights potential moons in the residuals, see Fig.~\ref{fig:deconv}f. Finally, the deconvolved images were derotated by the parallactic angle into sky-coordinate. 

We wish to draw the reader's attention to the high dynamics at play in this method. The reduced data peak to $4500$ Analog to Digital Units (ADUs), while the halo spans on the full dynamic down to a few tens of ADUs. After the deconvolution by the extended PSF which is recovered across three orders of magnitude, the photometry of the asteroid jumps close to $~10000$ ADU while S/2026 (44) 1 hides within less than 10 ADUs dynamic. \KM{This point-like source can be followed throughout the full sequence of the two nights (UT 2026-02-15 and 2026-03-21).}

The resulting video from the LBT deconvolutions can be found in Visualization 1. Some individual LBT frames and the VLT observations can be found in \KM{Fig.~\ref{fig:shape1}}. 

\subsection{Principal Component Analysis}
\label{sec:app_PCA}

The SHARK-VIS data are pupil stabilized: the instrument PSF orientation is fixed with respect to the detector while the field rotates with the parallactic angle. This allowed us to adapt the Angular Differential Imaging method classically used for exoplanet detection \citep[ADI,][]{2006ApJ...641..556M}, in order to confirm the detection of the satellite with an independent approach. From the co-aligned frames of Sect.~\ref{sec:app_co_align} we computed an ADI based on Principal Component
Analysis \cite[PCA-ADI,][]{2012MNRAS.427..948A, 2012ApJ...755L..28S}.
It consists of subtracting from each frame its projection on a limited number of PCA bases (first 1000 components), then de-rotating the frames, and finally computing their average stack. Due to the fast orbital motion of the putative moon, we split these resulting frames in batches of 10000 before average stacking, thus obtaining a series of sequential stacks (40 batches for the first night, and 11 for the second night).

In sixteen of these resulting stacks for UT 2026-02-15 and three stacks for UT 2026-03-21, we \KM{confirmed the detection of} a point-like source
about 215 mas and 175 mas respectively from Nysa’s center \KM{as seen in the blind deconvolution residuals}. \KM{These stacks were consecutive during the portion of the night with the best seeing and therefore best AO performance.} Figure~\ref{fig:moonPCA}, which is the maximum value stack of these selected stacks, highlights the moon orbital motion which is clearly visible.

\onecolumn
\section{Supplementary tables}

This appendix section contains additional tables describing the imaging datasets and photometric lightcurves.

\begin{table}[h!]
\caption{\label{tab:sphere} Observing geometry during LBT/SHARK-VIS and VLT/SPHERE/ZIMPOL observations of (44) Nysa}
\centering
\begin{tabular}{lccc}
\hline\hline
Date&UT&$\Delta$ (au)&Phase angle ($^o$)\\
\hline
2026-02-15* & &1.14&12.7\\
2026-03-09 &00:21&1.30&21.6\\
2026-03-11 &01:58&1.32&22.3\\
2026-03-15 &03:33&1.36&23.4\\
2026-03-16 &03:33&1.37&23.7\\
2026-03-21* & &1.43&24.8\\
2026-03-22 &00:47&1.44&25.0\\
2026-03-22 &02:15&1.44&25.0\\
2026-03-24 &02:38&1.45&25.4\\
2026-04-01 &01:23&1.54&26.7\\
2026-04-01 &02:30&1.54&26.7\\
2026-04-05 &01:30&1.59&27.1\\
\hline
\end{tabular}
\tablefoot{Dates and observational geometry of Nysa during the imaging dataset. Dates marked with a * are SHARK-VIS observations, all others are SPHERE. Listed time marks the UT time during the first exposure. Observing geometry accessed from \url{https://ssp.imcce.fr/forms/ephemeris}}
\end{table}

\begin{table}[h!]
    \centering
    \caption{Observing parameters for the LBT SHARK-VIS observations}
    \begin{tabular}{cccccc}
        \hline\hline\\
         UT & Filters & $\lambda$ &Width (nm) & 
           \multicolumn{2}{c}{Exposure time} \\
         &&&& Total (hours) & Per frame (ms) \\
         \hline
         2026-02-15T04:20 to 07:44 &  R Bessel & 645&140 & 3.3 & 25 \\
         2026-03-21T02:34 to 05:57 &  R Bessel & 645&140 & 1.6 & 25 \\
           &  V Bessel & 533&77  & 1.6 & 80 \\
         \hline
         
    \end{tabular}
    \label{tab:shark}

\end{table}

\begin{table}[h!]
\centering
    \caption[Photometric lightcurves of (44) Nysa.]{
      Date, duration ($\mathcal{L}$, in hours), number of points ($\mathcal{N}_p$), phase angle ($\alpha$),
      filter(s), and observatory
      for each
      lightcurve. \label{tab:lc}}
      \begin{tabular}{crrrll}
        \hline\hline
        Date & \multicolumn{1}{c}{$\mathcal{L}$} & \multicolumn{1}{c}{$\mathcal{N}_p$} &
        \multicolumn{1}{c}{$\alpha$} & Filt. &\multicolumn{1}{c}{Obs.} \\
        & \multicolumn{1}{c}{(h)} && \multicolumn{1}{c}{(\degr)}& &\\
        \hline
    
        2026-02-22 & 7.5 & 823  & 16.1  & R & Robinson  \\
        2026-03-02 & 7.5 &  46 &  19.4  & B,V,R,I  & Command Module\\
        2026-03-04 &  6.3 &  90 &   20.1  &  B,V,R,I & Command Module \\
        2026-03-05 &  5.3 &  69 &   20.4 & B,V,R,I & Command Module \\
        2026-03-05 & 5.1 & 114 & 20.4 &   R,B  &  Robinson \\
        2026-03-06 &  5.1 & 187 & 21.0 &   R,B  &  Robinson  \\
        2026-03-07 &  6.6 & 257 & 21.3 &   R,B  &  Robinson \\
        2026-03-10 &  2.7 & 123 & 22.0 &   L  &  Robinson  \\
        2026-03-14 &  4.3 &  292 &   23.3     & RC & TRAPPIST-N\\
        2026-03-15 &  4.2 &  198 &   23.6   & CN & TRAPPIST-N\\
        2026-03-16 &  3.3 &  147 &   23.8   & C2 & TRAPPIST-N\\
        2026-05-19 &  2.9 &  267 &  27.3    & R & TRAPPIST-S\\
        2026-05-22 &  0.3 &  28 &  27.0   & R & TRAPPIST-S\\
        2026-05-24 &  2.7 &  248 & 26.9    & R & TRAPPIST-S\\
        2026-05-25 &  1.7 &  144 & 26.8    & R & TRAPPIST-S\\
        \hline
        \end{tabular}
  \end{table}


\FloatBarrier
\section{Shape model}

\begin{figure*}[h!]
    \centering
    \includegraphics[width=0.33\linewidth]{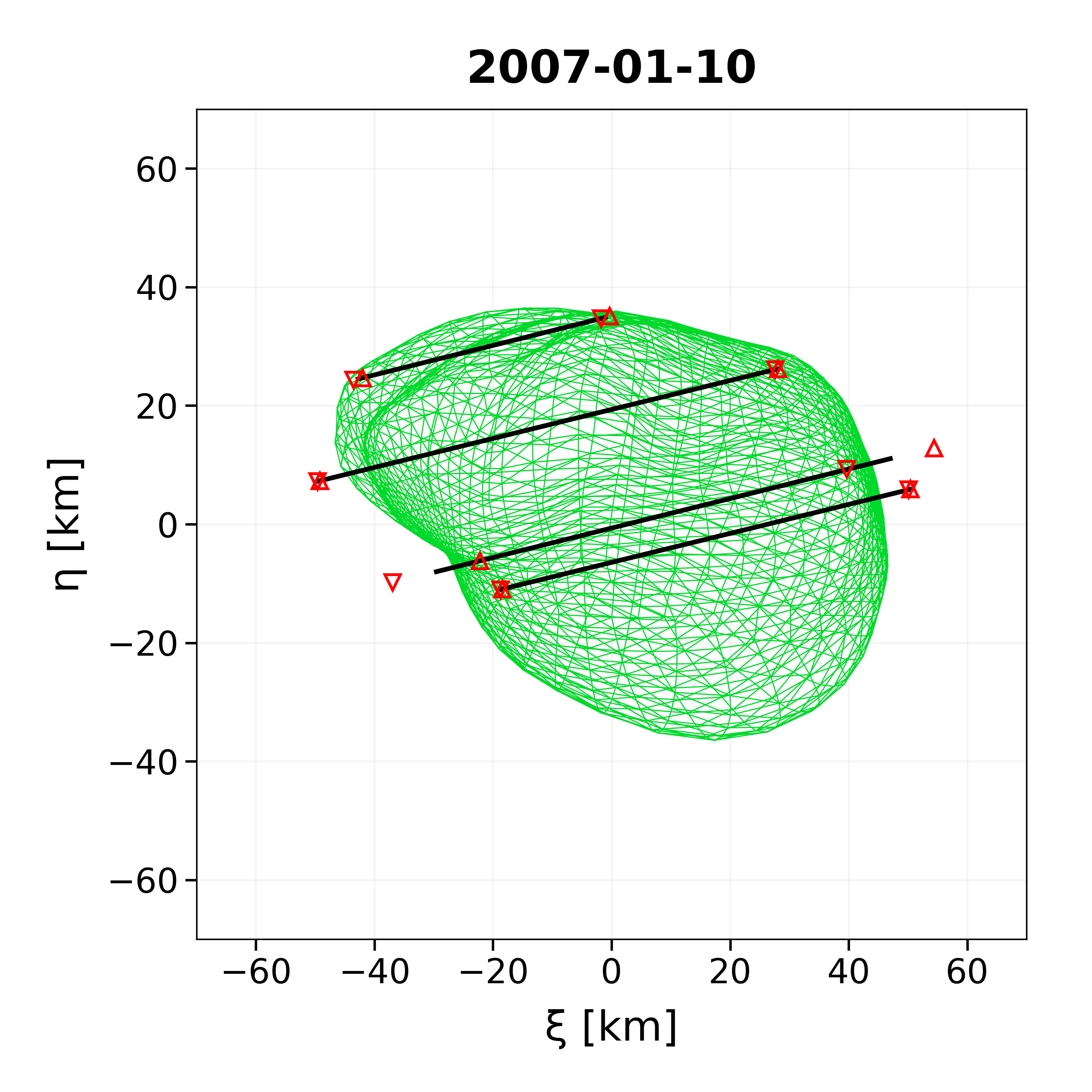}
    \includegraphics[width=0.33\linewidth]{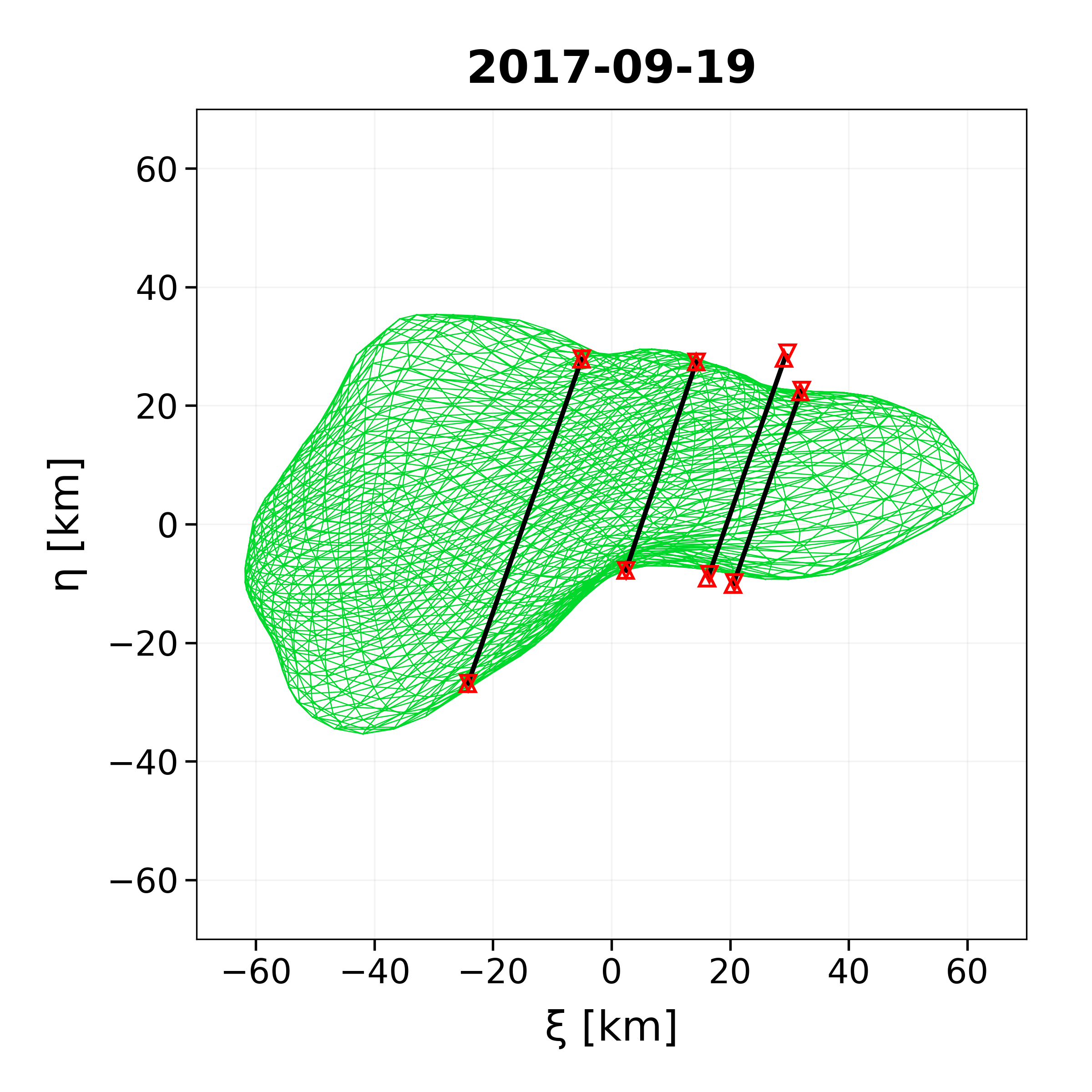}
    \includegraphics[width=0.33\linewidth]{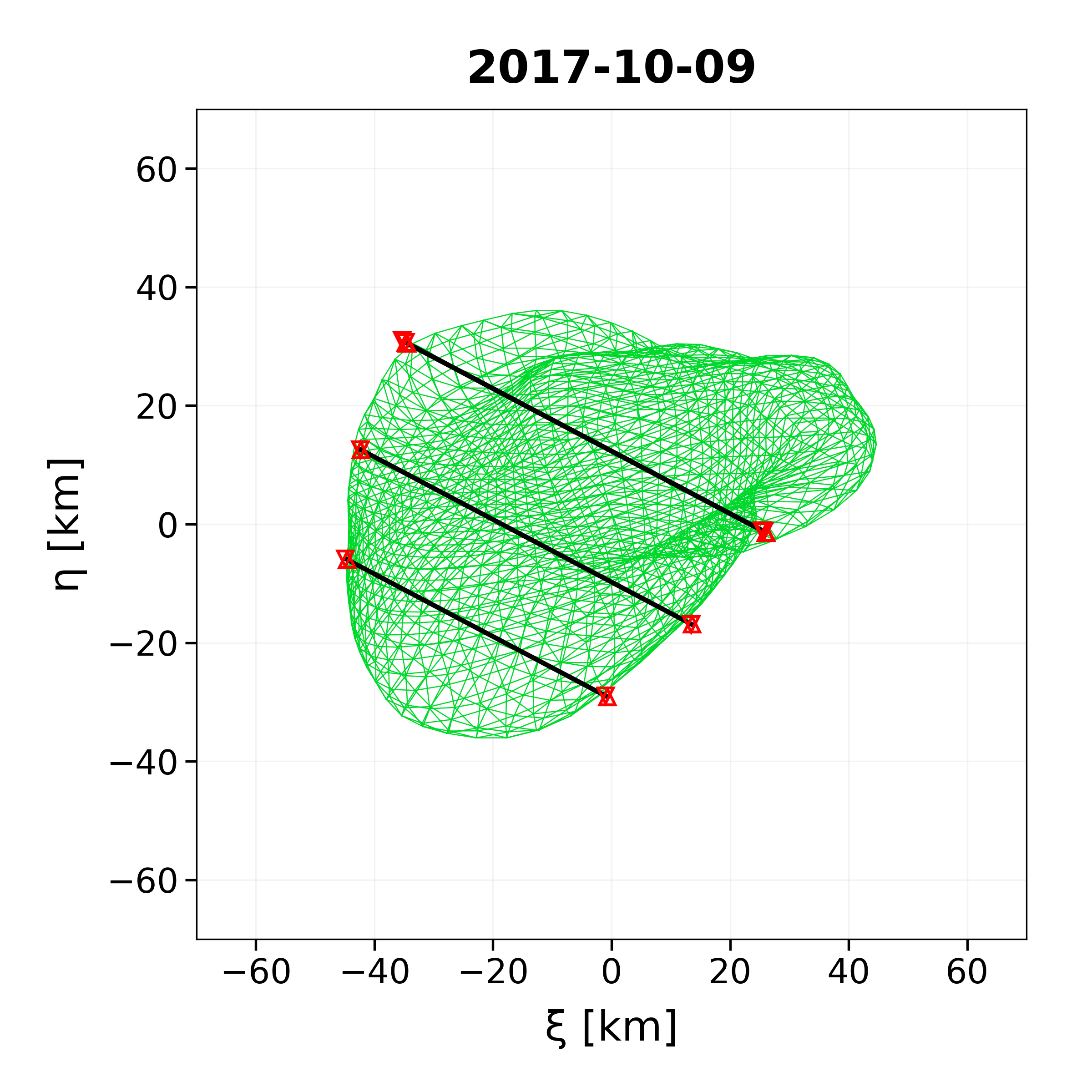}
    \caption{
    Observed occultation chords (black line segments) for three stellar occultations of Nysa, each with at least four positive chords, projected onto the sky plane, together with the final non-convex ADAM shape model (green mesh) at the corresponding rotational phases. The comparison is shown without any additional fitting. The close agreement confirms the validity of our size, shape, and spin-state solution.}
    \label{fig:occ}
\end{figure*}

\begin{figure*}[ht!]
\centering
\includegraphics[width=0.9\linewidth]{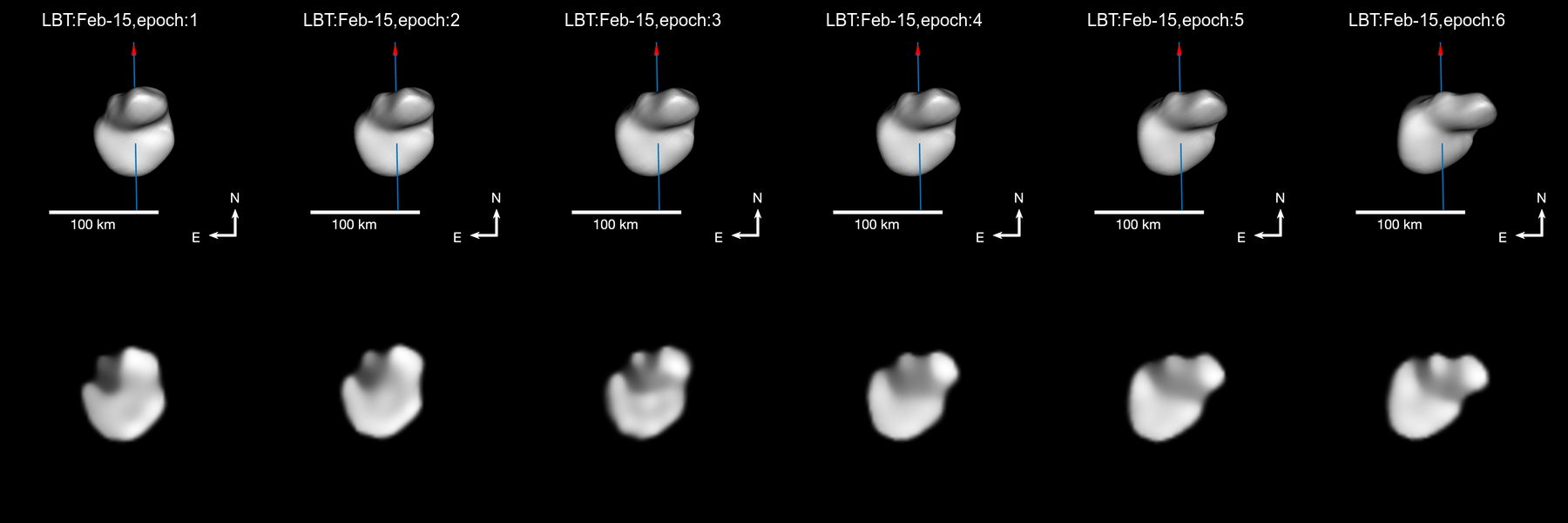}
\includegraphics[width=0.9\linewidth]{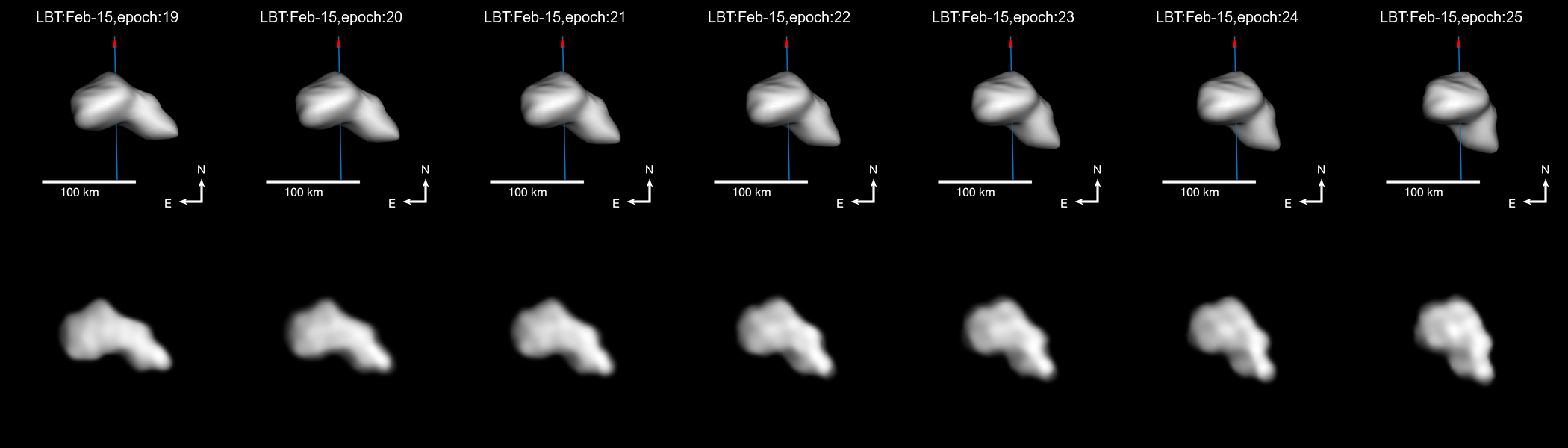}
\includegraphics[width=0.9\linewidth]{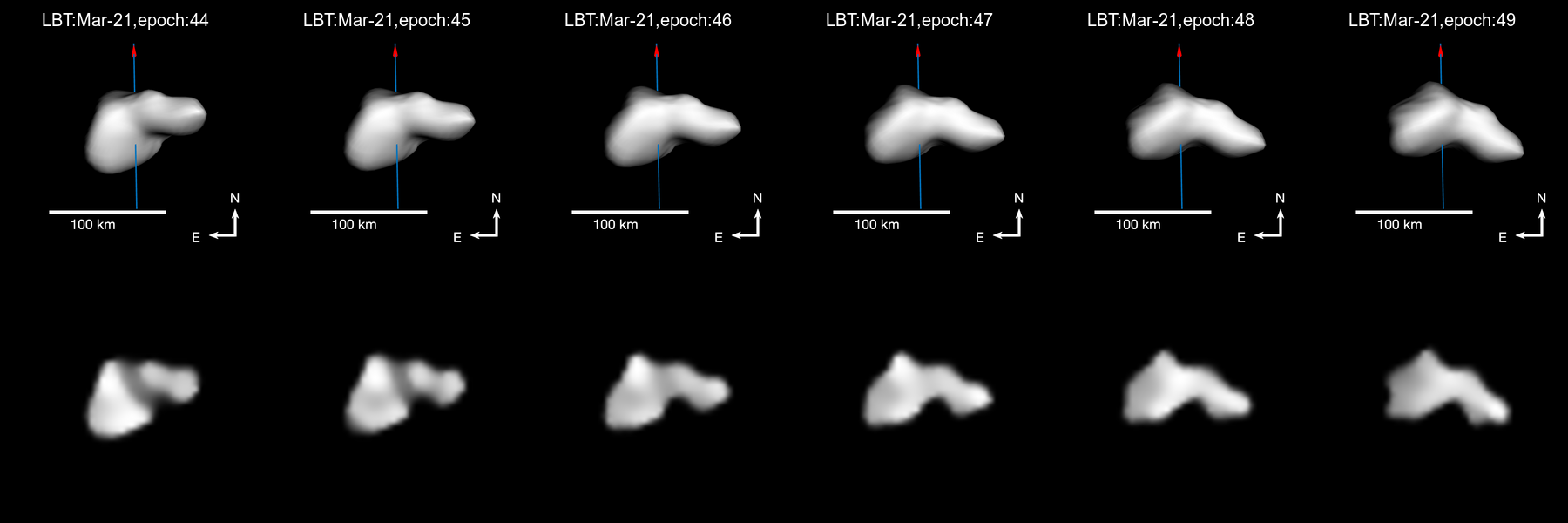}
\includegraphics[width=0.9\linewidth]{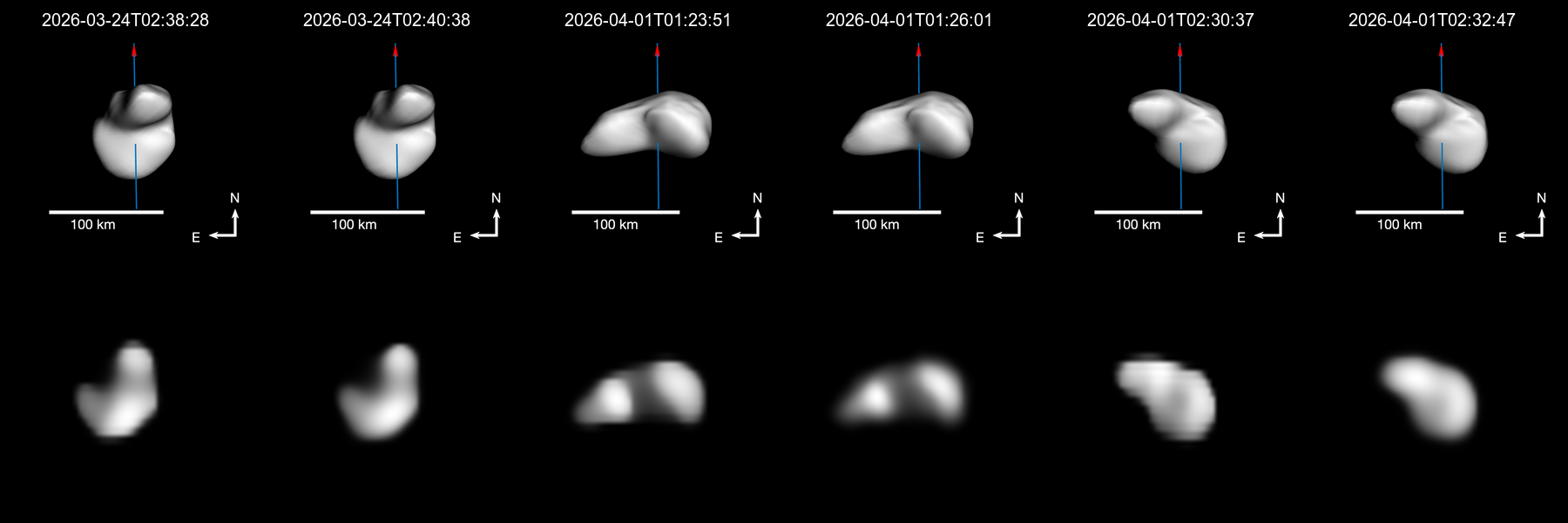}
\caption{Comparison between the deconvolved images of (44) Nysa (lower rows) and the corresponding projections of the ADAM shape model (upper rows). The red arrows indicate the direction of the spin axis. \KM{The full dataset is available on DAMIT \citep{2010A&A...513A..46D}.}}
\label{fig:shape1}
\end{figure*}

\begin{figure*}
\centering
\foreach \i in {1,...,15}{%
  \includegraphics[width=0.30\textwidth]{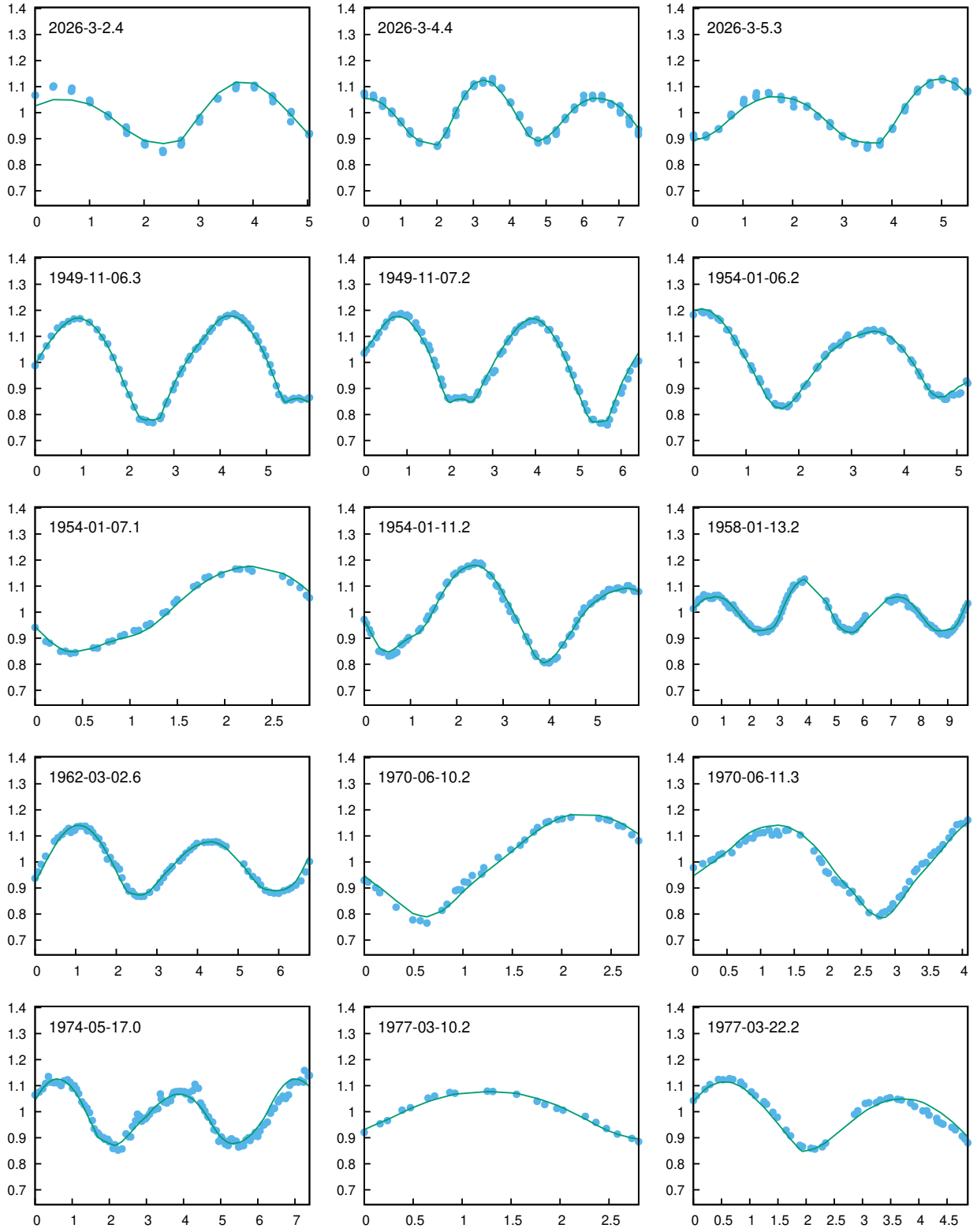}%
  \ifnum\i=3\\\fi%
  \ifnum\i=6\\\fi%
  \ifnum\i=9\\\fi%
  \ifnum\i=12\\\fi%
}
\caption{Comparison between the observed and synthetic lightcurves of asteroid (44) Nysa. Blue points correspond to disk-integrated photometric measurements, while green curves represent synthetic lightcurves computed from the final ADAM shape model. The horizontal axis shows the time from the beginning of each observing run in hours, and the vertical axis shows the relative intensity normalized to the mean brightness of the corresponding lightcurve. The observing epoch is indicated in the upper-left corner of each panel. \KM{The full sample of 83 lightcurve fits is available on DAMIT \citep{2010A&A...513A..46D}.}}
\label{fig:nysa_lcs_1}
\end{figure*}

\KM{This section contains information about the fit of Nysa's shape model as described in Sec. \ref{sec:shape}. Figure \ref{fig:occ} shows the fit of the shape model presented in Sec. \ref{sec:shape} compared to historical occultation data. Information about the date, location, and observers of each occultation can be found in Tab. \ref{tab:nysa_occultations}. Figure \ref{fig:shape1} shows the fit of the model to the imaging dataset described in Sec. \ref{sec:img}, and Fig. \ref{fig:nysa_lcs_1} shows the fit of the model to the photometric light curves described in Sec. \ref{sec:lc}. Figures \ref{fig:shape1} and \ref{fig:nysa_lcs_1} are representative of a more extenive dataset which can be found on DAMIT\footnote{\url{https://damit.cuni.cz/projects/damit/}} \citep{2010A&A...513A..46D}.. The shape model can also be downloaded at this location.}

\begin{table}
\caption{\KM{Stellar occultation observations of asteroid (44)~Nysa with more than three chords.}}
\label{tab:nysa_occultations}
\centering
\setlength{\tabcolsep}{3.0pt}

\begin{tabular}{@{}lp{3.2cm}rrrlrlrc}
\hline\hline
Observer & Location & $\lambda$ ($^\circ$) & $\phi$ ($^\circ$) & $h$ (m) &
$t_{\rm D}$ (UTC) & $\sigma_{\rm D}$ (s) &
$t_{\rm R}$ (UTC) & $\sigma_{\rm R}$ (s) & Type\\
\hline

\multicolumn{10}{c}{\textbf{2007 January 10}}\\
\hline
R.~Venable             & Clarks Hill, USA          & $-82.1876$ & $+33.6677$ & 128  & 01:30:47.3  & --   & --          & --   & M\\
P.~Maley               & Liberty City, USA         & $-94.9535$ & $+32.4094$ & 148  & 01:32:31.5  & --   & 01:32:37.0  & --   & P\\
R.~Venable             & Harlem, USA               & $-82.3203$ & $+33.4572$ & 140  & 01:30:43.00 & 0.03 & 01:30:53.11 & 0.03 & P\\
R.~Nugent              & Jacksonville, USA         & $-95.1576$ & $+32.0615$ & 100  & 01:32:28.8  & 1.00 & 01:32:38.9  & 1.00 & P\\
R.~Venable             & Wrens, USA                & $-82.3811$ & $+33.2240$ & 137  & 01:30:43.91 & 0.02 & 01:30:52.92 & 0.02 & P\\

\hline
\multicolumn{10}{c}{\textbf{2017 September 19}}\\
\hline
J.~Rovira              & Spain                     & $-3.1619$  & $+40.6052$ & 753  & 03:28:52.78 & 0.04 & 03:29:25.16 & 0.06 & P\\
F.~Casarramona         & Spain                     & $-2.0153$  & $+36.8806$ & 50   & 03:32:42.19 & 0.03 & 03:33:02.96 & 0.02 & P\\
J.~Ripero              & Spain                     & $-3.5806$  & $+40.6429$ & 648  & 03:29:01.40 & 0.20 & 03:29:23.40 & 0.30 & P\\
P.~Maley               & Spain                     & $-3.5581$  & $+40.4485$ & 610  & 03:29:11.89 & 0.10 & 03:29:31.10 & 0.10 & P\\
F.~Aceituno            & Calar Alto, Spain         & $-3.3861$  & $+37.0629$ & 2896 & 03:32:18.7  & --   & --          & --   & M\\

\hline
\multicolumn{10}{c}{\textbf{2017 October 9}}\\
\hline
B.~Leonard             & Grandview, TX, USA        & $-97.1674$ & $+32.2706$ & 203  & 04:09:12.01 & 0.06 & 04:09:22.85 & 0.06 & P\\
M.~Smith               & Grandview, TX, USA        & $-97.1675$ & $+32.2706$ & 203  & 04:09:11.98 & 0.03 & 04:09:22.78 & 0.03 & P\\
J.~Barton, R.~Campbell & Clifton, TX, USA          & $-97.6713$ & $+31.6755$ & 298  & 04:09:23.63 & 0.01 & 04:09:33.60 & 0.01 & P\\
D.~Eisfeldt            & Waco, TX, USA             & $-97.2499$ & $+31.6291$ & 161  & 04:09:23.34 & 0.03 & 04:09:31.18 & 0.03 & P\\
R.~Nugent              & Dripping Springs, TX, USA & $-98.1660$ & $+30.1951$ & 399  & 04:09:54.1  & --   & --          & --   & M\\
\hline
\end{tabular}

\tablefoot{The longitude $\lambda$ and latitude $\phi$ are geodetic coordinates, and $h$ is the altitude above sea level. The disappearance and reappearance times are denoted by $t_{\rm D}$ and $t_{\rm R}$, respectively, and their reported uncertainties by $\sigma_{\rm D}$ and $\sigma_{\rm R}$. The observation type is P for a positive detection and M for a miss. A dash indicates that no value was reported.}
\end{table}

\FloatBarrier

\clearpage
\section{Surface features}

\begin{figure*}[h!]
    \centering
    \includegraphics[width=1\linewidth]{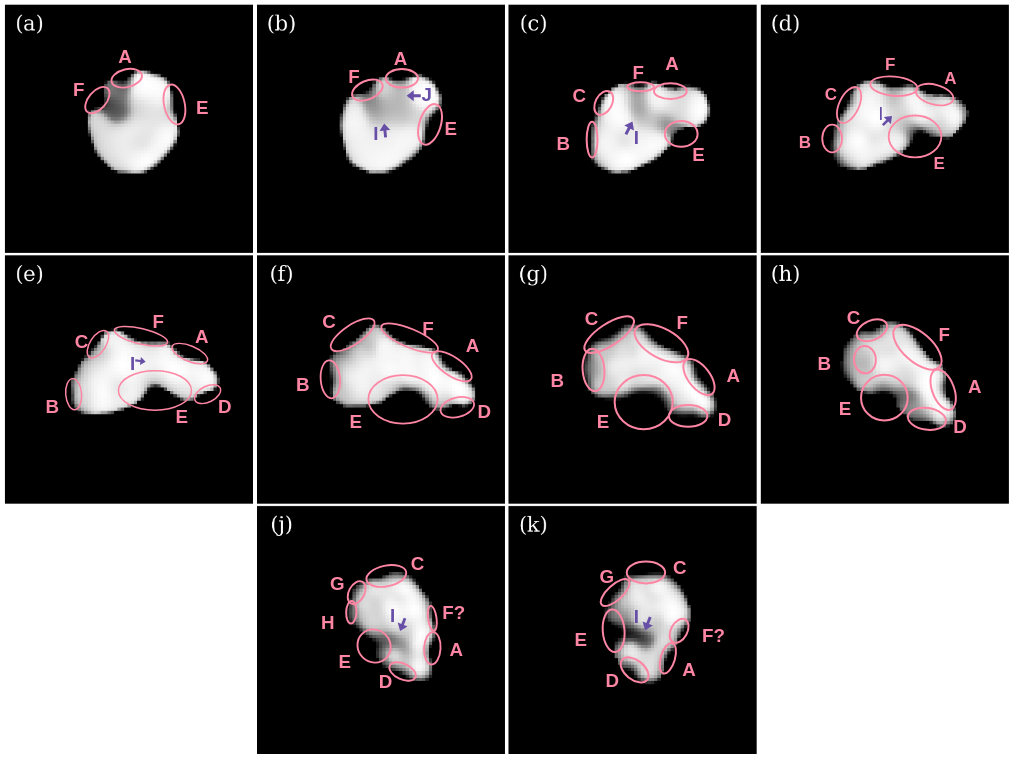}
    \caption{Surface features identified on Nysa in the 2026-02-15 SHARK-VIS dataset. Edge indentations are marked with pink ovals and shadowed features are marked with purple arrows.  Each feature is marked with a corresponding letter; the measured size and description of these features can found in table \ref{tab:craters}.}
    \label{fig:Nysacraters}
\end{figure*}

\begin{table}[h!]
\caption{\label{tab:craters} Surface features of Nysa.
}
\centering
\begin{tabular}{lccc}
\hline\hline
Feature&Type&Width (km)& Depth(km)\\
\hline
A          &Crater/secondary basin& 38 & 6\\
B &Crater     &20\\
C         &Crater&35\\
D     &Crater&21 & 3\\
E      &Southern cavity *  &53 & 17 \\
F      &Northern cavity  &44 & 5\\
G      &Crater&17\\
H      &Crater&8\\
I      &Collum/Valley   &\\
J      &Collum/Valley   & \\
\hline
\end{tabular}
\tablefoot{Crater width and depth correspond to the maximum measured projected width and depth of a feature. Reported measurements correspond to the frame in which the largest cross-section of the feature was visible. Uncertainties for all reported values are one pixel, or 5\,km. Depths are not reported for features with a depth less than two pixels. Indentations associated with a collum are denoted as basins, independent indentations are denoted as craters. Location of the features can be seen in Fig. \ref{fig:Nysacraters} and Fig. \ref{fig:contactriple}. 

* Ambiguous edge}
\end{table}

\onecolumn

\end{appendix}
\end{document}